\documentclass[11pt,fullpage]{article}
\usepackage{latexsym}
\usepackage{amssymb}
\usepackage{amsmath}
\usepackage{mathtools}
\usepackage{graphicx}
\usepackage{tikz}
\usetikzlibrary{arrows,shapes,positioning,automata,trees}
\usepackage{listings}
\usepackage{lscape}
\usepackage{multirow}
\usepackage{longtable}
\usepackage{diagbox}
\usepackage{CJKutf8}

\newcommand{\hide}[1]{}

\newtheorem{definition}{Definition}[section]

\newtheorem{theorem}[definition]{Theorem}
\newtheorem{lemma}[definition]{Lemma}
\newtheorem{corollary}[definition]{Corollary}

\newcommand{\mathdef}[1]{\relax\ifmmode #1\else $#1$\fi}

\newcommand{\true}{\mathdef{\mathit{true}}}

\renewcommand{\implies}{\rightarrow}
\newcommand{\ifonlyif}{\leftrightarrow}

\newcommand{\deq}{\stackrel{\scriptstyle\Delta}{=}}

\newcommand{\always}{\mathdef{\mbox{\raisebox{-.25ex}{$\Box$}}}}

\newcommand{\eventually}{\mathdef{\mbox{\raisebox{-.25ex}{$\Diamond$}}}}
\newcommand{\waitforop}{\relax\ifmmode \ \mathcal{W}\, \else $\mathcal{W}$ \fi}

\newcommand{\untilop}{\relax\ifmmode \ \mathcal{U}\, \else $\mathcal{U}$ \fi}

\newcommand{\releaseop}{\relax\ifmmode \ \mathcal{R}\, \else $\mathcal{R}$ \fi}

\newcommand{\mreleaseop}{\relax\ifmmode \ \mathcal{M}\, \else $\mathcal{M}$ \fi}

\newcommand{\sinceop}{\ifmmode \ \mathcal{S}\, \else $\mathcal{S}$ \fi}

\newcommand{\backtoop}{\ifmmode \ \mathcal{B}\, \else $\mathcal{B}$ \fi}

\newcommand{\triggerop}{\ifmmode \ \mathcal{T}\, \else $\mathcal{T}$ \fi}

\newcommand{\Uop}{\relax\ifmmode \ \textbf{U}\, \else $\textbf{U}$ \fi}

\newcommand{\Rop}{\relax\ifmmode \ \textbf{R}\, \else $\textbf{R}$ \fi}

\begin{document}

\title{From Ramsey-Based to Congruence-Based Constructions 
for B\"uchi Complementation
}

\author{Yih-Kuen Tsay\\
National Taiwan University
\and
Moshe Y. Vardi\\
Rice University
}
\date{September 5, 2026}

\maketitle

\begin{abstract}
The very first construction by J.~Richard~B\"uchi himself for complementing a B\"uchi 
automaton relies on a fundamental lemma about the division of an arbitrary infinite word 
into consecutive finite words, which was cleanly proven by invoking a specialized theorem 
of Ramsey.
For that reason, constructions of similar nature have subsequently been labeled as 
Ramsey-based.
Nevertheless, it suffices to have a weaker form of the lemma where the finite words come
from a finite number of congruence classes, rather than arbitrary classes, that 
form a partition of the set of all finite words.
The weaker lemma, with support of nicer properties from a congruence, can be proven 
without Ramsey's theorem.
A commonly adopted improvement on such complementation constructions also
requires the working of a congruence.
This paper recounts the history and reviews using more contemporary terminology
wherever possible the relevant concepts and results, to advocate renaming of 
Ramsey-based constructions as congruence-based constructions.
\end{abstract}

\tableofcontents

\section{Introduction}
\label{sec:introduction}
In his seminal 1962 paper~\cite{buchi:decision} where the essence of a formalism later 
called nondeterministic B\"uchi automata was conceived, J.~Richard~B\"uchi elucidated
the definability of infinitary languages in a restricted second-order theory of natural numbers, 
which he called 
the \emph{sequential calculus} (SC), and also affirmed the decidability of SC.
Today S1S (monadic second-order theory with one successor) is a more commonly used 
name for B\"uchi's SC, though the syntax is slightly different.
Subsequently, we will adopt the new name and yet follow the original syntax of SC
wherever more suitable for our purpose.

Formulae in S1S use, in addition to the usual first-order ingredients, 
monadic predicate variables ranging over subsets of natural numbers 
or tuples of such variables (just a convenient bundling of several related predicates),
and existential and universal quantifiers for such variables.
With monadic predicate variables or tuples of them, S1S formulae may be used to
describe constraints on an infinite word, which is an infinite sequence of symbols 
representing truth values of the monadic predicates when evaluated for a given
natural number.
Therefore, an S1S formula $A(w)$, with a free tuple $w$ of predicates 
(representing infinite words), defines some infinitary language.

Based on this observation, one key idea in the work of B\"uchi was to consider a 
particular class, named $\Sigma_1^\omega$, of S1S formulae. 
Let $\exists^\omega t(\ldots)$ be an abbreviation of 
$\forall n (\exists t(n<t \wedge \ldots)$ (what says that there are infinitely many 
natural numbers for $t$ s.t.\ $\ldots$).
A formula $B(w)$, with a free tuple $w$ of monadic predicates,  in the class 
$\Sigma_1^\omega$ has the following form:

\[
\exists r (I(r(0)) \wedge \forall t (\delta(r(t), w(t), r(t+1))) \wedge \exists^\omega t (F(r(t))))
\]

\noindent
where $I(r(0))$, $\delta(r(t), w(t), r(t+1))$, and $F(r(t))$ are propositional formulae
(the names of the formulae have been purposefully altered from those in the original 
presentation).

The $\Sigma_1^\omega$ formula $B(w)$, in effect, asserts the existence of an 
accepting run of some
(later-called) nondeterministic B\"uchi automaton on the input word $w$ and therefore 
defines some (later-called) $\omega$-regular language.
B\"uchi showed that every formula $A(w)$ in S1S can be effectively transformed
to an equivalent formula $B(w)$ in $\Sigma_1^\omega$.
It follows that formulae of the form $A(w)$ in S1S define exactly the class of 
$\omega$-regular languages and, utilizing the automaton-like structure in 
$\Sigma_1^\omega$ formulae that lead to the definability result, one can obtain
a decision procedure for the validity of any S1S sentence.

While extensively borrowing concepts and results from the theory of 
finite automata and regular languages, to obtain the needed closure of 
$\Sigma_1^\omega$ formulae (or languages defined by nondeterministic B\"uchi 
automata) under Boolean operations, particularly 
under complementation, B\"uchi uncovered
and made use of a fundamental lemma, called the \emph{sequential lemma}, 
about the division of an arbitrary infinite word into consecutive finite words.
More precisely, suppose sets $C_0, C_1, \ldots, C_n$ form a partition of the set
of all finite words over some given finite set $\Sigma$ of symbols.
The sequential lemma states that, for every infinite word $w$ over $\Sigma$, 
there exist $C_i$ and $C_j$ s.t.\ $w\in C_iC_j^\omega$, i.e., $w$ can be divided
as the concatenation of a word from $C_i$ and infinitely many words from $C_j$.
The sequential lemma was cleanly proven by invoking a specialized theorem of 
Ramsey.
For that reason, complementation constructions of similar nature have subsequently 
been labeled as Ramsey-based, e.g., ~\cite{fogarty:buchi,tsai:state,breuers:improved,allred:simple}.%
\footnote{We are unable to pinpoint where the term ``Ramsey-based construction'',
or the like, first appeared, but certainly it has been widely used in the literature.
It was probably derived from ``... The proof is based on Ramsey's Theorem ...'' in 
Sistla~\textit{et~al.}~\cite{sistla:complementation} or
 ``... a complementation construction 
that involved a Ramsey-based combinatorial argument ...'' in Vardi~\cite{vardi:buchi}.}

Nevertheless, it suffices to have a weaker form of the lemma where the finite words come
from a finite number of congruence classes, rather than arbitrary classes, that 
form a partition of the set of all finite words.
The weaker lemma, with support of nicer properties from a congruence, can be proven 
without Ramsey's theorem.
A commonly adopted improvement on such complementation constructions also
requires the working of a congruence.
This paper recounts the history and reviews using more contemporary terminology
wherever possible the relevant concepts and results, to advocate renaming of 
Ramsey-based constructions as congruence-based constructions.

\section{Preliminaries}
\label{sec:preliminaries}

\subsection{Sequences, Words, and Languages}
\label{subsec:sequences}
A \emph{sequence} over a finite set is an enumerated collection, which may be
finite or infinite, of elements from the set.
An input word to an automaton is a sequence and so is a run of
the automaton on the input word.
Besides conventionally using alpha-numeric names to identify the symbols 
in a word (such as $a$ or $b$) and the automaton locations in a run
(such as $l_0$ or $l_1$), we will also use elements from a set like 
$\{\emptyset,  \{p\}, \{q\}, \{p,q\}\}$, $\{\neg p\neg q,  p\neg q, \neg pq, pq\}$, 
$\{(\mathtt{F},\mathtt{F}), (\mathtt{T},\mathtt{F}), (\mathtt{F},\mathtt{T}), (\mathtt{T},\mathtt{T})\}$, or 
$\{\mathtt{FF}, \mathtt{TF}, \mathtt{FT}, \mathtt{TT}\}$ (representing all four truth 
assignments to a pair of propositions), so that desired conditions on a word or run
can be conveniently described by logical formulae, or in other words,
a word or run can serve as a model for interpreting logical formulae.

The \emph{length} of a sequence $x$ is the number of elements in $x$ and is denoted by $|x|$;
the length of the empty sequence $\varepsilon$ is $0$ and the length of an infinite sequence
is $\omega$.
$X^*$ is the set of all finite sequences over $X$ and
$X^\omega$ the set of all infinite sequences over $X$.
Given $x=x_0x_1\cdots x_{n-1} \in X^*$ and $y=y_0y_1y_2\cdots \in X^*\cup X^\omega$, 
$xy$ denotes the concatenation of $x$ and $y$, i.e., if $z=xy$, then $z_0=x_0$, $z_1=x_1$, \ldots,
$z_{n-1}=x_{n-1}$, $z_n=y_0$, $z_{n+1}=y_1$, and so on.
Concatenation is extended to two sets of sequences such that, for $Y\subseteq X^*$ and 
$Z\subseteq X^*\cup X^\omega$, $YZ$ denotes $\{yz\mid y\in Y\ \mbox{and}\ z\in Z\}$.

Given an infinite sequence $x=x_0x_1x_2\cdots\in X^\omega$, 
we write $x(i,j)$ to denote the
subsequence of $x$ from position $i$ to position $j-1$ (excluding $j$), 
namely $x_ix_{i+1}\cdots x_{j-1}$;
in the case of $i\geq j$, $x(i,j)$ is treated as the empty sequence.
And we write $x^i$ to denote the infinite subsequence, or 
suffix, of $x$ starting from $x_i$, i.e., $x^i_0=x_i$, $x^i_1=x_{i+1}$, $x^i_2=x_{i+2}$, and so on.

An infinite sequence over a finite set $X$ can also be seen as a mapping from the
set $\mathbb{N}$ of natural numbers (including $0$) to $X$.
That is, given a sequence $x=x_0x_1x_2\cdots\in X^\omega$, $x(0)=x_0$, $x(1)=x_1$, 
$x(2)=x_2$, and so on.
If $X$ is taken for example to be $\{\mathtt{FF}, \mathtt{TF}, \mathtt{FT}, \mathtt{TT}\}$,
then any $x\in X^\omega$ can be seen as a value assignment to a pair of 
monadic predicate variables ranging over subsets of natural numbers.
Alternatively, assuming $p$ and $q$ are given as the names of the two predicates, 
the set $X$ may be given as $\{\neg p\neg q,  p\neg q, \neg pq, pq\}$ for easier identification.
Such a set may serve as the alphabet for forming words and also
as the set of (binary-encoded) locations of an automaton.
Conditions can therefore be imposed on a word or automaton run by constraining
the corresponding monadic predicate or tuple of monadic predicates.

When the set $X$ is intended as the alphabet for forming words, we usually have the
name $\Sigma$ in place of $X$.
Each element in $\Sigma$ is called a \emph{symbol}.
A \emph{word} over $\Sigma$ is just another name for a sequence over $\Sigma$.
And so, $\Sigma^*$ is the set of all finite words over $\Sigma$ and
$\Sigma^\omega$ the set of all infinite words over $\Sigma$.
A \emph{finitary language} is a subset of $\Sigma^*$ and
an infinitary language is a subset of $\Sigma^\omega$.
We refer to either simply as \emph{language}, when no confusion may arise.

\subsection{B\"uchi Automata and $\omega$-Regular Languages}
\label{subsec:automata}
Nondeterministic B\"uchi (word) automata (NBWs) have the same structure as that of 
finite-state (word) automata but operate on infinite words~\cite{buchi:decision,thomas:automata,graedel:automata}.
For our purpose, we assume particular forms for the names of input symbols and  those of automaton locations.

\begin{definition}[Syntax of NBWs]
An NBW is a $5$-tuple $\langle\Sigma, Q, \delta, I, F\rangle$, where each component is as defined
below.
\begin{itemize}
\item $\Sigma$ is the finite \emph{alphabet}, of the form like $\{\emptyset,  \{p\}, \{q\}, \{p,q\}\}$, or
alternatively $\{\neg p\neg q,  p\neg q, \neg pq, pq\}$.
\item $Q$ is the finite set of \emph{locations}, of the form like $\{\mathtt{F}, \mathtt{T}\}$ or
$\{\mathtt{FF}, \mathtt{TF}, \mathtt{FT}, \mathtt{TT}\}$.
\item $\delta: Q\times\Sigma \rightarrow 2^Q$ is the \emph{transition function}.
  We also treat $\delta$ as a ternary \emph{relation}, whose members are called 
  \emph{transitions} (more about this below).
\item $I \subseteq Q$ is the set of initial locations.
We also treat $I$ as a predicate, writing $I(l)$ rather than $l\in I$, for $l\in Q$.
\item $F\subseteq Q$ is the set of acceptance locations.
We also treat $F$ as a predicate, writing $F(l)$ rather than $l\in F$, for $l\in Q$.
\end{itemize}
\end{definition}

When $l'\in\delta(l,a)$ holds for some $l,l'\in Q$ and $a\in\Sigma$, we also write it as $(l, a, l')\in\delta$ or $\delta(l,a,l')$ and call the triple $(l, a, l')$ a \emph{transition} and  $l'$ a \emph{successor} location of $l$.
This is extended to sets of locations and arbitrary finite words in the following sense.
For $l\in Q$, $Q'\subseteq Q$, $a\in \Sigma$, and $x\in \Sigma^*$, 
$\delta(l,\varepsilon)=\{l\}$, $\delta(l,ax)=\delta(\delta(l,a),x)$, and
$\delta(Q',x) = \bigcup_{l\in Q'} \delta(l,x)$.

An infinite word as the input drives an NBW to go in every step from one location to zero,
one, or several other locations, either failing to continue or successfully producing (infinite)
runs.
An NBW accepts an infinite word if there exists a run of the NBW on the word that visits
some location infinitely many times.

\begin{definition}[Semantics of NBW]
For an NBW $A=\langle\Sigma, Q, \delta, I, F\rangle$,
a run of $A$ on an infinite word $w=w_0w_1w_2\cdots\in\Sigma^{\omega}$
is an infinite sequence of locations $l_0l_1l_2\cdots\in Q^{\omega}$
such that $l_0\in I$ and for every $i \geq 0$, $(l_i,w_i,l_{i+1}) \in \delta$.
Given a run $\rho$, let $\mathrm{inf}(\rho)$ denote the set of locations that appear
infinitely many times in $\rho$.
A run $\rho$ on a word is \emph{accepting} if $\mathrm{inf}(\rho) \cap F \neq \emptyset$.
An input word $w$ is \emph{accepted} by an NBW $A$ if there is an accepting
run of $A$ on $w$.
\end{definition}

The set of words accepted by $A$ is called the language \emph{recognized} by $A$, 
or simply the \emph{language of $A$}, or denoted $L(A)$.

\begin{figure}
\begin{center}
\begin{tikzpicture}[
  auto,
  initial text=,
  semithick,
  >=stealth
]
  \node[state,initial,accepting](q0) at (2.4,-0.88){$\mathtt{F}$};
  \node[state](q1) at (5.6,-0.88){$\mathtt{T}$};
  \path[->]
    (q0) edge [loop above] node {$\neg p$,\ $q$} ()
    (q0) edge [bend left=10] node {$p\neg q$} (q1)
    (q1) edge [bend left=10] node {$q$} (q0)
    (q1) edge [loop above] node {$\neg q$} ()
  ;
\end{tikzpicture}
\end{center}
\caption{An NBW equivalent to $\always(p \implies \eventually q)$ (``$p$ leads to $q$'').
The single $q$ as a label on the transition edge from $\mathtt{F}$ to itself or from $\mathtt{T}$ 
to $\mathtt{F}$
is a shorthand for $p q$, $\neg p q$, i.e., $\{p, q\}, \{q\}$. 
So, either edge actually represents two
transitions, one labeled with  $p q$ (i.e., $\{p, q\}$) and the other with $\neg p q$ (i.e., $\{q\}$);
analogously for the $\neg p$-labeled and $\neg q$-labeled edges.}
\label{fig:nbw_example}
\end{figure}
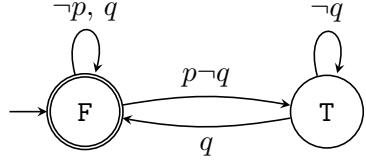

An automaton and a logical formula are \emph{equivalent} when they specify the
same language.
In Figure~\ref{fig:nbw_example} is an NBW that is equivalent to the LTL formula
$\always(p \implies \eventually q)$ (``$p$ leads to $q$''), asserting that every (positive) occurrence of $p$ 
is accompanied at the same position or followed later by an occurrence of $q$.
In the diagram, an initial location is indicated by an incoming arrow without a label, 
while a location in the acceptance set is double-circled.
The language of the NBW may also be defined using a $\Sigma_1^\omega$ formula
$\exists r (I(r(0)) \wedge \forall t (\delta(r(t), w(t), r(t+1))) \wedge \exists^\omega t (F(r(t))))$
with the following subformulae:
\begin{itemize}
\item $I(r(0)) \deq \neg r(0)$
\item $\delta(r(t),w(t),r(t+1))$ $\deq$
         $\begin{array}{ll}
         & (\neg r(t)\wedge \neg p(w(t)) \wedge \neg r(t+1))\\
         \vee & (\neg r(t)\wedge q(w(t)) \wedge \neg r(t+1) )\\
         \vee & (\neg r(t)\wedge p(w(t)) \wedge \neg q(w(t)) \wedge r(t+1))\\
         \vee & (r(t)\wedge \neg q(w(t)) \wedge (r(t+1)))\\
         \vee & (r(t)\wedge q(w(t)) \wedge \neg r(t+1))
         \end{array}$
         
where $p(w(t))$ is \true\ iff $w(t)=p\neg q$ or $p q$ and $q(w(t))$ is \true\ iff $w(t)=\neg p q$ 
or $p q$.
\item $F(r(t)) \deq \neg r(t)$
\end{itemize}

A language $L\subseteq\Sigma^\omega$ is \emph{$\omega$-regular} if
$L=\bigcup_{i=1}^{n}U_iV_i^\omega$, where $U_i,V_i\subseteq\Sigma^+$ are regular.
The following statement are equivalent~\cite{buchi:decision,thomas:automata}:
\begin{itemize}
\item $L$ is $\omega$-regular.
\item $L$ is recognized by some NBW.
\item $L$ is defined by some $\Sigma_1^\omega$ formula.
\end{itemize}

The class of $\omega$-regular languages is effectively closed under Boolean operations.
A proof of this property was one of the main contributions of B\"uchi's work, where
the case of closure under complementation is rather hard.
B\"uchi's original proof dealt with $\Sigma_1^\omega$ formulae.
We dive here into the difficult case of complementation, but reformulate the proof
for NBWs.
Many other later formalisms exist for defining the class of $\omega$-regular 
languages, for some of which closure under complementation 
is trivial (with the price of making some other thing hard)~\cite{mcnaughton:testing,graedel:automata}.

\section{Sequential Lemma and Universal Coverage}

\begin{lemma}[Sequential Lemma~\cite{buchi:decision}; Universal Coverage]
Let $\Sigma$ be a finite alphabet and
suppose $C_0, C_1, \ldots, C_n$ form a partition of $\Sigma^*$.
Then, for every infinite word $w\in\Sigma^\omega$, there exist $C_i$ and $C_j$,  
where $0\leq i,j\leq n$, s.t.\ $w\in C_iC_j^\omega$, i.e.,
$\Sigma^\omega=\bigcup\{C_iC_j^\omega\mid 0\leq i,j\leq n\}$.
\end{lemma}

The lemma says that, given an arbitrary $w=a_0a_1a_2\ldots a_{i-1}a_i\ldots\in\Sigma^\omega$, 
it is always possible to divide $w$ into an infinite number of finite word segments 
$a_0\ldots a_{i_0-1} \mid a_{i_0}\ldots a_{i_1-1} \mid a_{i_1}\ldots a_{i_2-1} \mid \ldots \mid 
a_{i_{n-1}}\ldots a_{i_n - 1} \mid \ldots$
such
that all the word segments except probably the first belong to the same $C_i$, where
$0\leq i\leq n$.

It can be proven by using Ramsey's Theorem, whose original presentation is recreated 
below for easy reference.
A specialized form of the theorem suffices and we provide two equivalent presentations 
to follow the original theorem.

\begin{theorem}[\textsc{Theorem A} of Ramsey~\cite{ramsey:problem}]
Let $\Gamma$ be an infinite class, and $\mu$ and $r$ positive integers;
and let all those sub-classes of $\Gamma$ which have exactly $r$ members,
or, as we may say, let all $r$-combinations of the members of $\Gamma$ be
divided in any manner into $\mu$ mutually exclusive classes $C_i$
($i=1, 2, \ldots, \mu$), so that every $r$-combination is a member of one and 
only one $C_i$; then, assuming the axiom of selections, $\Gamma$ must
contain an infinite sub-class $\Delta$ such that all the $r$-combinations of
the members of $\Delta$ belong to the same $C_i$.
\end{theorem}

\hide{
\footnotesize
[Ramsey 1930] Ramsey, F.P., ``On a Problem of Formal Logic,'' 
\textit{Proc. London Math. Soc.}, Series 2, Volume 30, 1930, pp. 264--286.
(Also in: Gessel, I. and Rota, G.C. (eds), \textit{Classic Papers in Combinatorics}, 
2009. Modern Birkhäuser Classics. Birkhäuser Boston.)
}

\begin{theorem}[Specialized Theorem A of Ramsey]
Let $G$ be an infinite complete undirected graph and $c$ a positive integer;
and let each edge of $G$ be colored with one of $c$ different colors,
then, $G$ must contain an infinite monochromatic clique.
\end{theorem}

\begin{theorem}[Specialized Theorem A of Ramsey]
\label{theorem:specialramsey}
Let $\mathbb{N}$ be the set of natural numbers and
$k$ a positive integer;
and let all pairs $\{m,n\}$, where $m,n\in\mathbb{N}$ and
$m\neq n$, be divided into $k$ mutually
exclusive sets $C_i$ ($i=1, 2, \ldots, k$),
then there is an infinite subset of $\mathbb{N}$ such that all
pairs of distinct numbers in the subset belong to the same $C_i$.
\end{theorem}

Below is a proof of the sequential lemma~\cite{buchi:decision,sistla:complementation}.

\begin{figure}
\begin{center}
\scriptsize
\begin{tikzpicture}[
  x=1cm, y=1cm,
  auto,
  state/.style={draw,circle,minimum size=.25cm},
  elliptic state/.style={draw,ellipse,minimum width=1cm,minimum height=.25cm},
  initial text=,
  semithick,
  >=stealth,
]
  \node[state](s00) at (0.0,0.0){$0$};
  \node[state](s01) at (1.0,0.0){$1$};
  \node[state](s02) at (2.0,0.0){$2$};
  \node[state](s03) at (3.0,0.0){$3$};
  \node[state](s04) at (4.0,0.0){$4$};
  \node[state](s05) at (5.0,0.0){$5$};
  \node[state](s06) at (6.0,0.0){$6$};
  \node[state](s07) at (7.0,0.0){$7$};
  \node[state](s08) at (8.0,0.0){$8$};
  \node[state](s09) at (9.0,0.0){$9$};
  \node[state,draw=none](s09a) at (10.0,0.0){$\cdots$};

  \path
    (s01) edge [bend left=30] node {2} (s06)
 ;
\end{tikzpicture}
\end{center}
\caption{Coloring. If $w(1,6)=a_1a_2\cdots a_5$ belongs to Class 2, then the undirected edge
$\{1,6\}$ is colored with Color 2.}
\label{fig:coloring}
\end{figure}
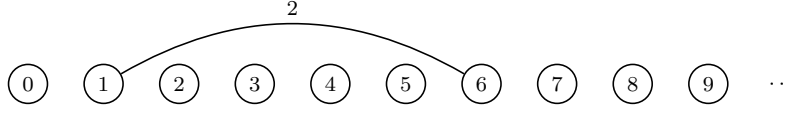

Suppose $w=a_0a_1a_2\ldots a_{i-1}a_i\ldots\in\Sigma^\omega$.
Recall that $w(i,j)$, where $i<j$, denotes the non-empty finite word segment $a_ia_{i+1}\cdots a_{j-1}$ 
of $w$.
So, every pair of distinct natural numbers corresponds to a non-empty finite word segment of $w$
and may be assigned to the corresponding set in the partition, 
as illustrated by Figure~\ref{fig:coloring}.

From Specialized Theorem A of Ramsey (Theorem~\ref{theorem:specialramsey}), 
there exists an infinite subset of natural numbers
such that all pairs of distinct numbers in the subset belong to the same $C_i$
(are of the same color).
Suppose the subset is $\{i_0, i_1, i_2, \ldots, i_n, \ldots\}$, where $i_0<i_1<i_2<\cdots<i_n<\dots$.
($i_0$ may be equal to $0$).
Then, $w(i_0,i_1)=a_{i_0}\cdots a_{i_1-1}$, $w(i_1,i_2)=a_{i_1}\ldots a_{i_2-1}$, $\ldots$, 
$w(i_{n-1},i_n)=a_{i_{n-1}}\ldots a_{i_n-1}$, etc.\ all belong to the same $C_i$.

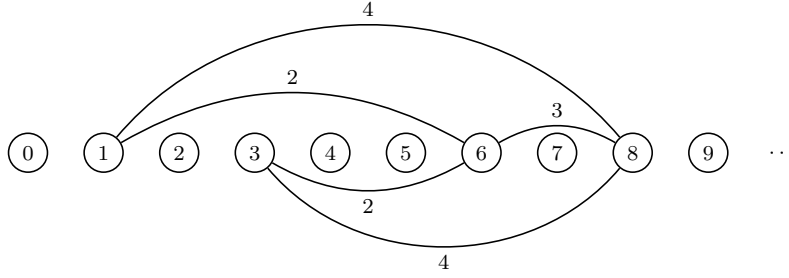
\begin{figure}
\begin{center}
\scriptsize
\begin{tikzpicture}[
  x=1cm, y=1cm,
  auto,
  state/.style={draw,circle,minimum size=.25cm},
  elliptic state/.style={draw,ellipse,minimum width=1cm,minimum height=.25cm},
  initial text=,
  semithick,
  >=stealth,
]
  \node[state](s00) at (0.0,0.0){$0$};
  \node[state](s01) at (1.0,0.0){$1$};
  \node[state](s02) at (2.0,0.0){$2$};
  \node[state](s03) at (3.0,0.0){$3$};
  \node[state](s04) at (4.0,0.0){$4$};
  \node[state](s05) at (5.0,0.0){$5$};
  \node[state](s06) at (6.0,0.0){$6$};
  \node[state](s07) at (7.0,0.0){$7$};
  \node[state](s08) at (8.0,0.0){$8$};
  \node[state](s09) at (9.0,0.0){$9$};
  \node[state,draw=none](s09a) at (10.0,0.0){$\cdots$};

  \path
    (s01) edge [bend left=30] node {2} (s06)
    (s03) edge [bend right=30] node[swap] {2} (s06)
    (s06) edge [bend left=30] node {3} (s08)
    (s01) edge [bend left=50] node {4} (s08)
    (s03) edge [bend right=50] node[swap] {4} (s08)
 ;
\end{tikzpicture}
\end{center}
\caption{Congruence classes. If $w(1,6)\cong w(3,6)$, then $w(1,8)\cong w(3,8)$.
An arbitrary coloring need not have this property.}
\label{fig:congruence}
\end{figure}

The sequential lemma assumes nothing further about the partition.
If the partition is assumed to be formed by the congruence classes of a
congruence over $\Sigma^*$ of finite index, one can certainly draw the same 
conclusion.
A congruence has nicer properties than an arbitrary partition or coloring, 
as illustrated by Figure~\ref{fig:congruence}, but
invoking Ramey's Theorem A does not fully exploit the properties of 
a congruence of finite index.
With the further assumption about the partition, a simpler and more direct proof 
is possible.

A \textit{congruence} $\cong$ over $\Sigma^*$ is an equivalence relation
compatible with concatenation, i.e., satisfying:
\[
\forall x,y,y',z\in\Sigma^*\ (y \cong y'\ \ \Rightarrow\ \ xyz\cong xy'z).
\]

Denote by $[u]_{\cong}$, or simply $[u]$, the $\cong$-class of $u$ restricted to 
$\Sigma^+$.
A congruence is of \textit{finite index} if it has a finite number of classes.
If $\cong$ is a congruence over $\Sigma^*$ of finite index, then, for every $u\in\Sigma^*$, 
$[u]\subseteq\Sigma^+$ is regular.

\begin{lemma}[Universal Coverage by a Congruence~\cite{sistla:complementation,thomas:automata}]
\label{lemma:uccong}
Let $\Sigma$ be a finite alphabet and 
suppose $\cong$ is a congruence over $\Sigma^*$ of finite index.
Then, for every infinite word $w\in\Sigma^\omega$,
there exist $[u]$ and $[v]$ in $\cong$-classes s.t.\ $w\in [u][v]^\omega$, i.e.,
$\Sigma^\omega=
\bigcup\{[u][v]^\omega\mid [u],[v]\in \mbox{$\cong$-classes}\}$.
\end{lemma}

Let $\cong$ be as in Lemma~\ref{lemma:uccong} and
$w=a_0a_1a_2\ldots a_{i-1}a_i\ldots\in\Sigma^\omega$.
Two positions (indices) $j$ and $j'$ ($j<j'$) are said to \textit{merge} at a
later position $m$ ($j'<m$) if $w(j,m) \cong w(j',m)$~\cite{mcnaughton:testing,thomas:automata}.
Clearly, if $j$ and $j'$ merge at $m$, then they merge also at every $m'>m$.
In Figure~\ref{fig:congruence}, for example, positions $1$ and $3$ merge at position $6$ and
also at a later position $8$.
We write $j \equiv_w j'$ if $j$ and $j'$ merge at some $m$.
For a given $w\in\Sigma^\omega$, $\equiv_w$ is an equivalence relation of 
finite index, since $\cong$ is of finite index.

Below is a proof of Lemma~\ref{lemma:uccong} 
(the weaker sequential lemma) without Ramsey's theorem, following
essentially the one given in~\cite{thomas:automata}.

Given an arbitrary $w=a_0a_1a_2\ldots a_{i-1}a_i\ldots\in\Sigma^\omega$.
There is an infinite subsequence $j_0, j_1, j_2, ...$ (of the positions) 
in which every pair of positions merge, since the merge relation $\equiv_w$ is of finite index.
One may assume $0 < j_0$ (or, we simply drop the leading $0$).
Further, there is an infinite subsequence $k_0, k_1, k_2, \ldots$ ($0<j_0\leq k_0$) of 
the above subsequence such that $w(k_0,k_1) \cong w(k_0,k_2) \cong w(k_0,k_3) \cong ...$,
since $\cong$ is of finite index.
To sum up, we now have an infinite sequence $(0<)\ k_0<k_1<k_2<k_3<\cdots$ such that
  \begin{itemize}
  \item every pair of $k_0, k_1, k_2, k_3, \ldots$ merge, 
  \item $w(0,k_0)\in [u]$, for some $u\in \Sigma^*$, and
  \item $w(k_0,k_1), w(k_0,k_2), w(k_0,k_3), \ldots \in [v]$, for some $v\in \Sigma^*$.
  \end{itemize}

Every pair of $k_0, k_1, k_2, \ldots$ merge and a pair of positions merge at $m$ 
also merge at every $m'>m$.
So, we may further select an infinite subsequence $(k_0=)\ l_0, l_1, l_2, \ldots$ such that
  \begin{itemize}
  \item every pair of $l_0, l_1, \ldots, l_i$ merge at $l_{i+1}$ for every $i>0$ and
  \item $w(l_0,l_i) \in [v]$ for every $i>0$, for some $v\in \Sigma^*$.
  \end{itemize}

We now have $w(l_i,l_{i+1})\in [v]$ for every $i\geq 0$, since, for $i=0$, $w(l_0,l_1)\in [v]$ and,
for every $i>0$, $w(l_i,l_{i+1})\cong w(l_0,l_{i+1}) \in [v]$ 
(from that $l_0$ and $l_i$ merge at $l_{i+1}$).
This completes the proof.

A pair $([u],[v])$ of congruence classes is said to be \emph{proper} if
$[u][v]\subseteq [u]$ and $[v][v]\subseteq [v]$.
Lemma~\ref{lemma:uccong} still holds even if restricted to proper pairs of congruence 
classes.
In its proof above, 
since $w(l_0,l_i)$, $w(l_i,l_{i+1})$, and $w(l_0,l_{i+1})=w(l_0,l_i)w(l_i,l_{i+1})$ are all in $[v]$, 
we have $[v][v]\cap [v]\neq \emptyset$, which implies $[v][v]\subseteq [v]$.
Also, if we started out with an infinite subsequence $0<j_1<j_2<\cdots<j_n<\cdots$, 
such that $w(0,j_1)$, $w(0,j_2)$, $\ldots$ $w(0,j_n)$, $\ldots$, etc. all belong to 
the same class, say $[u]$, then we have $[u][v]\subseteq [u]$.

\begin{lemma}[Proper Universal Coverage~\cite{sistla:complementation}]
Let $\Sigma$ be a finite alphabet and 
suppose $\cong$ is a congruence over $\Sigma^*$ of finite index.

Then, for every $w\in\Sigma^\omega$,
there exists a proper pair $([u],[v])$ of $\cong$-classes s.t.\ $w\in [u][v]^\omega$, i.e.,
$\Sigma^\omega=\bigcup\{[u][v]^\omega\mid \mbox{([u],[v]) is a proper pair of
$\cong$-classes}\}$.

\end{lemma}

\section{B\"uchi Complementation}
The B\"uchi complementation problem is to obtain, for a given NBW, 
another NBW whose language is the complement of that of the given NBW, which
proves the closure of the class of $\Sigma_1^\omega$ formulae under complementation.

Aside from the lemma for universal coverage by a congruence, we also need the notion of 
a congruence saturating a language.
A congruence $\cong$ is said to \textit{saturate} $L$ if, for every $\cong$-class $[u]$ and $[v]$,
$[u][v]^\omega\cap L\neq\emptyset$ $\Rightarrow$ $[u][v]^\omega\subseteq L$.
When $\cong$ \textit{saturates} $L$, then $[u][v]^\omega$ is either completely inside $L$
or completely outside $L$, i.e., inside $\overline{L}$.

\begin{theorem}[Language Coverage~\cite{buchi:decision,sistla:complementation}]
Let $\cong$ be a congruence over $\Sigma^*$ of finite index.
If $\cong$ saturates $L$,
then $L=\bigcup\{[u][v]^\omega\mid
uv^\omega\in L\}$.
\end{theorem}

And, together with the lemma of universal coverage by a congruence, this implies the 
following corollary.

\begin{corollary}[Complement Coverage~\cite{buchi:decision,sistla:complementation}]
\label{corollary:complement}
Let $\cong$ be a congruence over $\Sigma^*$ of finite index.
If $\cong$ saturates $L$,
then $\overline{L}=\bigcup\{[u][v]^\omega\mid uv^\omega\not\in L\}$.
\end{corollary}

Given an NBW $A=\langle\Sigma, Q, \delta, I, F\rangle$, we can define an equivalence
relation $\cong_A$, detailed below, over $\Sigma^*$ that is a congruence of finite index
saturating $L(A)$.
The $\cong_A$-classes are all regular and hence an NBW can be constructed for 
$[u][v]^\omega$, for any pair $([u],[v])$ of $\cong_A$-classes, and also for a finite union 
of languages of this form.
Corollary~\ref{corollary:complement} implies that, for a given NBW, 
another NBW whose language is the complement of that of the given NBW can be
constructed.

To define $\cong_A$, let us consider classifying finite words according to how they
drive the NBW $A$.
For $l,l'\in Q$ and $x\in\Sigma^*$, when $l'\in\delta(l,x)$ holds, we also write it
as $\delta(l,x,l')$.
If a run segment (a finite sequence of automaton locations) confirming 
the truth of $\delta(l,x,l')$ contains an acceptance location, then we write $\delta^F(l,x,l')$ to 
indicate so.

For $x,y\in\Sigma^*$,
\[
\begin{array}{l}
x \cong_A y\\
\mbox{iff}\\
\mbox{for every $l,l'\in Q$, $\delta(l,x,l')\ifonlyif \delta(l,y,l')$ and
$\delta^F(l,x,l')\ifonlyif \delta^F(l,y,l')$.}
\end{array}
\]

It can be shown that $\cong_A$ is a congruence over $\Sigma^*$ of finite index
and it saturates $L(A)$.
This congruence was used by B\"uchi \cite{buchi:decision} and also by 
Sistla~\textit{et~al.}~\cite{sistla:complementation} to complete their respective 
complementation construction.
B\"uchi worked with $\Sigma_1^\omega$ formulae and did not give a bound
on the size of the resulting complement $\Sigma_1^\omega$ formula.
In the latter work, the notion of a proper pair of 
$\cong_A$-classes was defined and used to reduce the size of 
the resulting complement automaton, though the asymptotic bound $2^{O(n^2)}$
remains the same.

\section{Remarks}
The reviewed complementation construction, based on a congruence, for NBWs 
does not really require the generality of the original sequential lemma, which may
be useful for other more complex classes of languages.
A weaker form of the lemma, where the needed partition of the set of all finite words 
is formed by congruence classes, suffices and it can be proven without
Ramsey's theorem.
Labeling such complementation constructions as Ramsey-based suggests 
the necessity of something highly sophisticated or even mysterious, which might
be a deterrent for them to be spread more widely.
One may actually argue that they are conceptually the simplest among all existing 
kinds of complementation constructions, 
including determinization~\cite{safra:complexity}, 
slice~\cite{kahler:complementation}, and rank-based~\cite{kupferman:weak} ones.
A more accurate and informative label clearly is congruence-based.

\begin{quote}
\upshape
\begin{CJK*}{UTF8}{bsmi}
$\cdots$

子曰：「必也正名乎！」

$\cdots$

子曰：「$\ldots$ 名不正，則言不順；$\ldots$」

\medskip
\hfill 
-- 《論語》子路第十三
\end{CJK*}
\end{quote}

In English:

\begin{quote}
\upshape
$\cdots$

The Master replied, ``What is necessary is to rectify names.''

$\cdots$

The Master said, ``$\ldots$
If names be not correct, language is not in accordance with the truth of things.
$\ldots$''

\medskip
\hfill 
-- \textit{The Analects (of Confucius), Chapter 13}
\end{quote}

\end{document}